\documentclass[aip,jcp,reprint]{revtex4-1}

\usepackage{graphicx}
\usepackage{dcolumn}
\usepackage{bm}

\usepackage[utf8]{inputenc}
\usepackage[T1]{fontenc}
\usepackage{mathptmx}
\usepackage{etoolbox}
\def\ph2{{\em p}-H$_2$}
\def\od2{{\em o}-D$_2$}

\makeatletter
\def\@email#1#2{%
 \endgroup
 \patchcmd{\titleblock@produce}
  {\frontmatter@RRAPformat}
  {\frontmatter@RRAPformat{\produce@RRAP{*#1\href{mailto:#2}{#2}}}\frontmatter@RRAPformat}
  {}{}
}%
\makeatother
\begin{document}

\preprint{AIP/123-QED}

\title{Isotopic separation in mixed clusters of molecular hydrogen}
\author{Kiril  M. Kolevski}
\author{Jie-Ru Hu}%
 \altaffiliation[Present address:]{State Key Laboratory of Precision Spectroscopy, East China Normal University, Shanghai 200062, China.}
\author{Massimo Boninsegni}
 \email{m.boninsegni@ualberta.ca.}
\affiliation{ 
Department of Physics, University of Alberta, Edmonton, Alberta, Canada T6G 2H5
}%

\date{\today}

\begin{abstract}
We investigate mixed (50/50) clusters of parahydrogen and orthodeuterium at low temperature, by means of Quantum Monte Carlo simulations. Our results provide evidence of liquid-like behavior and partial isotopic separation in a cluster of 640 molecules, at temperature $T=10$ K. As the temperature is lowered below $\sim 6$ K, crystallization occurs, with no indication that the liquid phase is more resilient at low temperature in a mixed cluster. Isotopic separation is therefore predicted to take place at low temperature only through the slow process of molecular self-diffusion in a crystalline matrix.
\end{abstract}

\maketitle

\section{\label{intro}Introduction}

A mixture of two atomic or molecular species of different masses undergoes isotopic separation at low temperature, in the presence of identical {\em inter}- and {\em intra}-species interactions. This is a remarkable manifestation of quantum mechanics; it would not take place classically, even at temperature $T=0$, due to the residual entropy of mixing. Quantum statistics usually plays an important role in isotopic separation; for example, in $^3$He-$^4$He mixtures the difference in quantum statistics results in the finite solubility of $^3$He in $^4$He in the $T\to 0$ limit, while $^4$He is excluded from $^3$He \cite{Graf67,Ebner71,Boninsegni95}. For the case of Bose statistics, the effective attraction arising from quantum exchanges results in the low-temperature separation of two species even when the masses are equal \cite{Jain2011,Jain2013}.
\\ \indent
However, isotopic separation takes place also in moderately quantum systems, in which quantum statistics is altogether negligible (e.g., rare gas solids and liquids \cite{Cuccoli93}). In this case, a difference in mass results in different bulk equilibrium densities for two isotopes, which at low temperature renders the demixed state energetically favorable. A physical system in which isotopic separation should in principle be observed is a mixture of parahydrogen (\ph2) and orthodeuterium (\od2) molecules, which, at ordinary conditions of pressure and temperature, can be regarded as ``elementary particles'', as electronic transfer and non-adiabatic effects are negligible. The interaction among two \ph2 (\od2) molecules, as well as between a \ph2 and a \od2, can all be described by the same potential, which to a good approximation can be considered spherically symmetric\cite{Operetto06,Operetto07}. Indeed, isotopic separation has been predicted \cite{Prigogine54,Gaines64,White65,Mullin69} for this system below a temperature $\lesssim 3$ K.
\\ \indent
Experimentally, however, no such separation is observed in bulk \ph2/\od2 mixtures\cite{Souers23}. The accepted explanation is that it is preempted by crystallization, occurring at a temperature between 14 and 19 K (depending on the relative isotopic concentration). Self-diffusion of molecules through a crystalline matrix is notoriously a very slow and inefficient process, rendering the observation of isotopic separation unfeasible in practice. Indeed, the tendency of molecular hydrogen to solidify at low temperature is very pronounced, even in reduced dimensions \cite{Boninsegni04,Boninsegni13}.  Now, this is also the conventional explanation for the inability to demonstrate experimentally the speculated superfluid transition of liquid molecular hydrogen, predicted many years ago to occur at temperature of a few K \cite{Ginzburg72}. Attempts to supercool liquid \ph2 down to a temperature where the superfluid transition may take place have been unsuccessful\cite{Bretz81,Seidel86,Sokol96,Schindler96}, but it is worth noting that theoretical studies based on realistic microscopic potentials have revised the estimated superfluid transition temperature to something $<$ 0.1 K  \cite{Boninsegni18}. Nonetheless, supercooling a fully mixed liquid mixture remains a plausible avenue for the observation of isotopic separation. 
\\ \indent
Meanwhile, there exists considerable theoretical \cite{Sindzingre91,Mezzacapo06,Mezzacapo07,Khairallah07,Mezzacapo09,Boninsegni20} and experimental \cite{Li10,Grebenev00,Kuyanov08} evidence of nanoscale size clusters of \ph2 remaining liquid at significantly lower temperature than the bulk, possibly displaying superfluid behavior (although it seems fair to say that definitive experimental confirmation is still lacking). It seems therefore compelling to look at clusters of isotopic mixtures of molecular hydrogen as a plausible experimental playground, to observe the quantum-mechanical effect of isotopic separation. Basic energy arguments suggest that, because the equilibrium density of the heavier isotope is greater\cite{Operetto06,Operetto07} (due to the lower zero-point kinetic energy), in the ground state of the system all heavier (\od2) molecules will be contained inside an inner spherical shell, surrounded by a shell of greater radius comprising all \ph2 molecules. The question, of course, is whether it is possible to observe such a separation at low temperature; a sufficiently large cluster is needed (e.g., of size of the order of a few hundred molecules), in order to make the whole notion of separation meaningful. The complication arises from the tendency of clusters of molecular hydrogen to crystallize (an assertion that we have independently verified in this work with a few targeted simulations) at temperatures as high as 6 K, for a number of molecules $\gtrsim 50$.
\\ \indent
In a recent article\cite{Sliter24}, the claim is made of experimental observation of such separation in nanometer-size mixed (50/50) clusters of \ph2 and \od2 supercooled down to a temperature as low as 2 K, at which clusters appear to be in a liquid-like state. This conclusion is supported by the results of an {\em ab initio} Molecular Dynamics simulation of the system in which ionic and electronic degrees of freedom are treated quantum-mechanically in a separate, inter-dependent way \cite{Deuk14}.
In this work, we carried out a different study, also based on computer simulations, but making use of imaginary-time equilibrium Quantum (Path Integral) Monte Carlo. The main differences between the calculation carried out in this work and that of Ref. \onlinecite{Sliter24} are the following:
\begin{enumerate}
    \item {We make use of a static, spherically symmetric pair potential to describe the interaction among hydrogen molecules, the same for all species. This has been the standard approach in virtually {\em all} studies of hydrogen clusters based on computer simulations. In particular, we utilize the standard 
    (Silvera-Goldman) intermolecular potential, which has been shown to provide a reasonably accurate, quantitative account of structural and energetic properties of the liquid and solid phases of \ph2, including the momentum distribution \cite{Boninsegni09,Hu23}. It is worth mentioning that all theoretical predictions of superfluidity in clusters of molecular hydrogen have been based on simulations making use of an intermolecular pair potential.}
    \item{Our simulation method is an {\em equilibrium} one, with no built-in time dependence. It is possible, however, to study sufficiently long-lived metastable phases on adopting a suitable cooling protocol\cite{Boninsegni18}. The key strength of this approach is that it is numerically {\em exact}, i.e., no uncontrolled approximations are introduced.}
\end{enumerate}
We consider mixed 50/50 \ph2/\od2 clusters comprising 640 molecules, initially prepared in a fully mixed state at a temperature $T=20$ K, and cool it down to 2.5 K, using different protocols; this simulated system closely matches that of the experiment and the simulation of Ref. \onlinecite{Sliter24}. 
\\ \indent
Our results show that at $T=10$ K both the mixed and a pristine \ph2 cluster are liquid-like, with partial but clearly noticeable separation of isotopes in the mixed cluster. 
As the temperature is lowered, the  propensity of the system to solidify emerges, and there is strong evidence that at $T=5$ K the cluster is in a crystalline configuration. As reported already in a previous study of the bulk phase \cite{Boninsegni21}, mixing the two isotopes does {\em not} result in any enhancement of the liquid phase and/or suppression of crystallization at low temperature. In fact, our results give compelling indication that the {\em opposite} is true, i.e., the orderly crystalline arrangement of molecules is favored by the mixing of the two isotopes. The separation of isotopes observed in our simulations at temperatures $\lesssim 5$ K invariably reflects what is already present in the liquid-like cluster at a higher (typically 10 K) temperature, where quantum effects of zero point motion are already quite important. As the cluster is cooled, molecules become essentially ``frozen'' in place, with the consequent dramatic slowdown of self-diffusion, i.e., the only physical mechanism that can result in isotopic separation.

\section{Methodology}
The mixture is described as an ensemble of $N=640$ point-like particles, of which $Nx$ are \ph2, the rest $N(1-x)$ \od2 molecules, both species obeying Bose statistics. We specifically consider the two cases $x=0.5$, i.e., a 50/50 mixture, as well as $x=1$, namely a pristine \ph2 cluster for comparison. The system is enclosed in a cubic cell with periodic boundary conditions in the three directions (but the side of the cell is 100 \AA, i.e., large enough to make boundary conditions immaterial). 
The quantum-mechanical many-body Hamiltonian of the system reads as follows:
\begin{eqnarray}\label{u}
\hat H = -\sum_{i\alpha}\lambda_{\alpha}\nabla^2_{i\alpha}+\sum_{i<j}v(r_{ij})
\end{eqnarray}
where the subscript $\alpha=H,D$ refers to either \ph2 ($H$) or \od2 ($D$). The first sum runs over all particles of either species, with $\lambda_{H}\ (\lambda_{D})=12.031 \ (6.0155)$ K\AA$^{2}$, whereas the second sum runs over all pairs of particles, $r_{ij}\equiv |{\bf r}_i-{\bf r}_j|$ and $v(r)$ is the Silvera-Goldman pair potential,\cite{Silvera78} which describes the interaction between two hydrogen molecules of either species. 
It is worth mentioning that, strictly speaking, \textit{inter}- and \textit{intra}- species interactions ought not be {\em identical}. On the contrary, one should expect some differences arising from the different nuclear masses. However, such differences are likely to be quantitatively small, and should not affect the main physical outcome of our study, especially at the temperatures of interest here. Furthermore, adopting just one interaction potential allows us to zero in on the effect of mass, which is believed to be at the root of isotopic separation.
\\ \indent
Because at temperatures as high as 20 K there is a significant tendency for molecules to evaporate off the cluster, we have included in our simulation a harmonic confining potential of length 10 \AA, in order to limit the loss of molecules to less than $\sim 5$\% at the highest temperature (evaporation is not a serious issue at temperatures less than 5 K).
\\ \indent
As repeatedly observed in previous studies of the bulk phase\cite{Boninsegni18} or clusters of size comparable to that considered here\cite{Boninsegni19}, exchanges are essentially non-existent in \ph2 (and {\em a fortiori} in \od2), at least at the temperatures considered here, as they are strongly suppressed by the relatively large size of the repulsive core of the intermolecular potential.
Therefore, we carried out simulations of the system at finite temperature by means of standard Path Integral Monte Carlo\cite{Boninsegni05}. In particular, we made use of an approximation for the short imaginary-time propagator which is accurate to fourth order in the time step $\tau$. Our results are extrapolated to the $\tau\to 0$ limit,  and we observed convergence of the estimate with a time step $\tau=3.125\times 10^{-3}$ K$^{-1}$. 
\\ \indent
Our investigation consists of cooling the cluster from a starting temperature of 20 K down to 2.5 K. Two different cooling protocols are adopted. In the first, the temperature of the equilibrated system is halved, first from 20\ K to 10\ K, then to 5 K and finally to 2.5\ K. In the second, a direct quench from 20\ K to 2\ K is carried out. We see no qualitative difference between the results in the two cases.

\section{Results and discussion}

\begin{figure}
\includegraphics[width=8.75 cm]{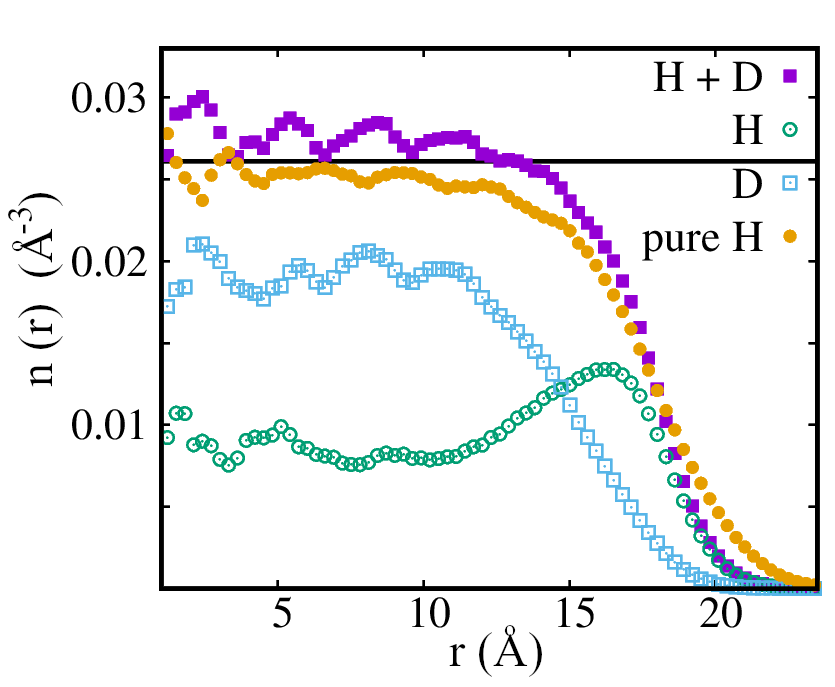}
\caption{Radial density profiles $n(r)$ for the two isotopes \ph2 (open circles) and \od2 (open boxes) for a 50/50 mixed cluster comprising altogether 640 molecules at a temperature $T=10$ K. Filled boxes show the cumulative density obtained by adding the results for the two isotopes. Also shown for comparison the corresponding profile computed for a pristine \ph2 cluster (filled circles). All profiles are  computed with respect to the center of the cluster. Solid horizontal line provides the \ph2 $T=0$ equilibrium density. }
  \label{T10rad}
\end{figure}
\begin{figure}
\includegraphics[width=8.75 cm]{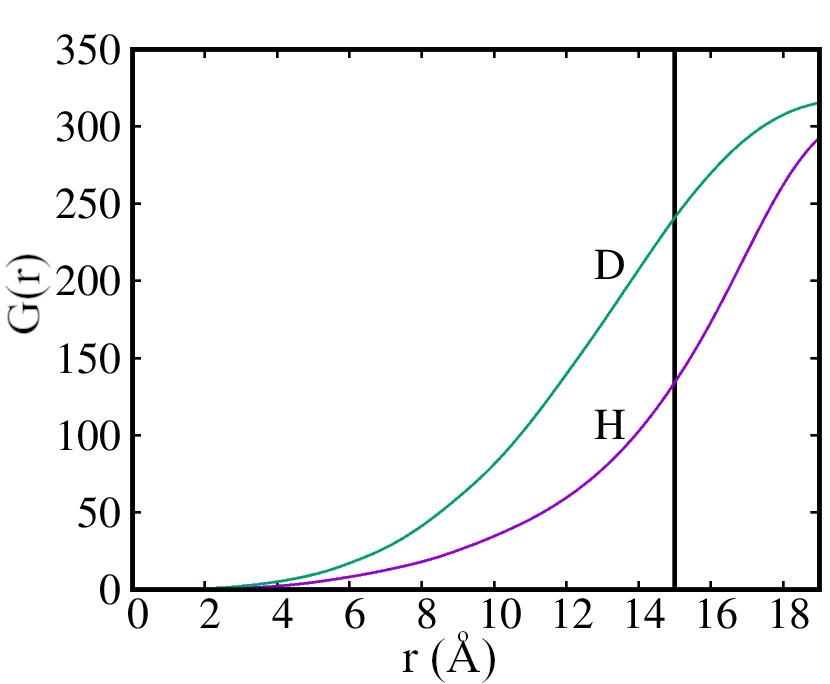}
\caption{Integrated density profiles $G(r)$ (see text), yielding the average numbers of molecules for each isotope inside a sphere of radius $r$ centered at the center of the cluster.}
\label{T10rad2}
\end{figure}
Fig. \ref{T10rad} shows the main result of a simulation of a mixed (50/50) cluster at temperature $T=10$ K. 
It shows (open symbols) the radial density profiles $n_\alpha(r)$, computed with respect to the center of the cluster. Also shown is the cumulative density profile (filled boxes), obtained by adding the profiles for the two isotopes, as well as the profile (filled circles) corresponding to a pristine \ph2 cluster with the same number of molecules (i.e., the $x=1$ limit of (\ref{u})). 
\\ \indent
The first observation is that none of the four density profiles displayed in Fig. \ref{T10rad} features the kind of sharp oscillations, signaling the presence of well-defined concentric shells, that typically characterize solid-like clusters. Instead, they are all relatively smooth and nearly ``flat'' in the central part, suggesting that both the mixed and the pristine cluster are in a liquid-like state, the mixed cluster being noticeably more compact and denser in its core. This in and of itself is not particularly surprising, as it is well known that a cluster of size like the one studied here will retain liquid-like features at temperatures lower than the freezing temperature of the bulk.
What seems remarkable, on the other hand, is that the mixed cluster shows clear evidence of isotopic separation, even at this relatively high temperature. This is apparent in the density profiles, with the \ph2 one featuring a shoulder at a distance $\sim 17$ \AA\ from the center of the cluster. \\ \indent 
This statement can be rendered more quantitative by looking at the integrated density profile $G_\alpha(r) = \int_{r^\prime < r} d^3r^\prime\ n_\alpha(r^\prime)$ (Fig. \ref{T10rad2}), yielding the number of particles of species $\alpha$ inside a sphere of radius $r$, centered at the center of the cluster. As shown in Fig. \ref{T10rad2}, a spherical shell of radius 15 \AA\ (vertical solid line) will contain approximately 75\% of all \od2 molecules, but less than 50\% of \ph2. \\
\begin{figure}
\includegraphics[width=8.75 cm]{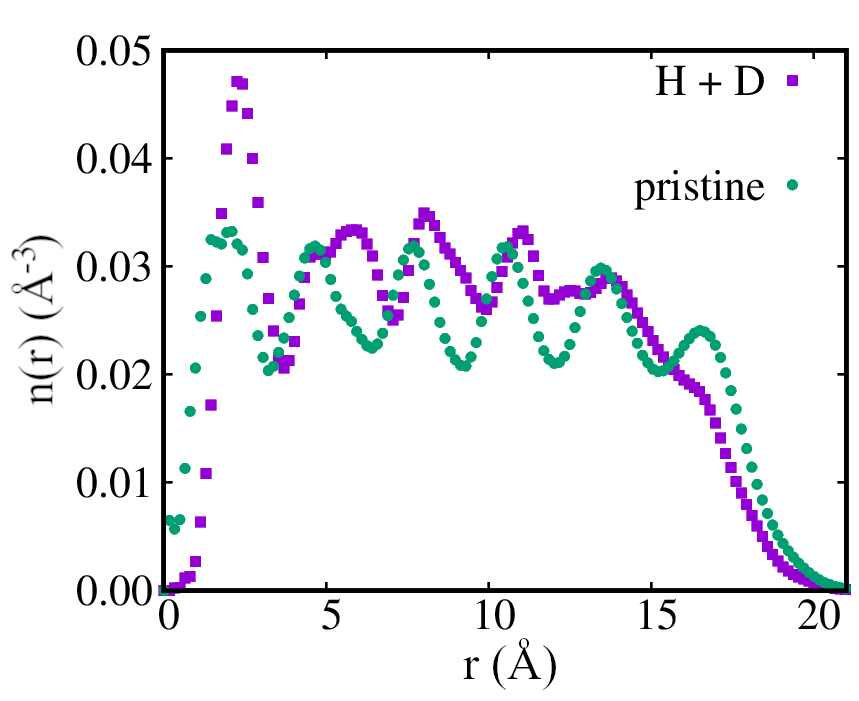}
\caption{Same as Fig. \ref{T10rad2} but for a temperature $T=2.5$ K.   The corresponding result for the integrated density $G(r)$ is virtually indistinguishable from that shown in the right part of Fig. \ref{T10rad}. The profile for the pristine cluster is obtained through a sufficiently long simulation, in which the cluster can be established to have reached thermal equilibrium  (as assessed based on standard simulation criteria). The mixed cluster profile is instead obtained through a quench of the cluster  at high ($T$=20 K) temperature; it should therefore be regarded as a ``typical'' result, as quenching from a different initial high temperature configuration will normally yield a slightly different profile (i.e., still with sharp peaks but at slightly different positions).}
  \label{T2p5rad}
\end{figure}
As previously mentioned, isotopic separation is driven by zero-point motion alone, as exchanges are absent in this system at these temperatures. The energetic contribution of zero-point kinetic energy, which is different for the two isotopes due to the different masses, drives the equilibrium density to higher values for \od2. Indeed, the kinetic energy in the mixed cluster is 52 K (46 K) for \ph2 (\od2) molecules, i.e., in both cases significantly higher than the classical (15 K) value. 
\\ \indent
Fig. \ref{T2p5rad} shows the result obtained at a temperature $T=2.5$ K. The result for the pristine cluster corresponds to an equilibrated system (i.e., a sufficiently long simulation), while for the mixed cluster we ``quenched'' the system from 20 K to the target temperature. What is shown in Fig.  \ref{T2p5rad} is a typical result. We also arrived at our target temperature for the mixed cluster by restaritng the simulation at $T=5$ K and halving the temperature, but we obtained the same qualitative result.

The density profiles for both the mixed and pristine cluster are clearly very different from those at $T=10$ K, i.e., they display marked oscillations, with sharp peaks, consistent with molecular localization. We repeated the quenching ``numerical experiment'' for the mixed cluster using different initial high temperature (i.e., 20 K) configurations; here it is important to note that, while the details of the final outcome (i.e., the positions of the peaks) may vary slightly on starting from different high temperature configurations, the most important physical feature, namely the appearance of sharp peaks, which we interpret as due to molecular localization, is a constant occurrence. We therefore conclude that it is a genuine physical effect, consistent with the crystallization of the mixed cluster.
\\ \indent
The mixed cluster is typically  more compressed than the pristine one, and the height of the peaks significantly higher. This is an indication that the mixing of the two isotopes {\em enhances} the tendency of the system to crystallize\cite{Boninsegni21}. We believe this to be due to the presence of heavier molecules which become localized at higher temperature and may act as nucleation centers. 
No evidence is obtained of greater demixing of the two isotopes at the lower temperature, with respect to what is seen at $T=10$ K. Indeed, the integrated density profiles at $T=2.5$ K are essentially indistinguishable from those shown on the right part of Fig. \ref{T10rad}. Again, we find this to be independent of our cooling protocol, and is consistent with increased molecular localization. We also mention that the results obtained at $T=5$ K are not quantitatively very different from those at $T=2.5$ K, i.e., strong evidence of crystallization is observed.

\section{Conclusions}
We have carried out Quantum Monte Carlo simulations of nanoscale size mixed \ph2/\od2 and pristine \ph2 clusters at low temperature, with the aim of providing an independent theoretical assessment of recently published work\cite{Sliter24}. Our computational methodology differs from that adopted therein in some important aspects. On the one hand, QMC allows us to obtain physical results that are numerically exact, statistical and systematic errors being reducible to a level where they do not significantly alter the physical conclusions. On the other hand, ours is an equilibrium approach, i.e., no time-dependent information can be extracted but only at best indirectly inferred.
Furthermore, the reliability of our study hinges on that of the intermolecular potentials, unlike the first principles simulations in Ref. \onlinecite{Sliter24} which, by contrast, are based on an intrinsically approximate framework (i.e., time-dependent Hartree).
The methodology developed in Ref. \onlinecite{Deuk14} has been shown to reproduce accurately some important experimental properties, but it is unclear whether it can provide a quantitative account of the energetics of the ions, which is the crucial physical mechanism underlying isotopic separation.
\\ \indent
Our results support some of the conclusions of Ref. \onlinecite{Sliter24}, as they point to isotopic separation in the liquid phase of the mixed clusters; however, we find such separation to be quantitatively more pronounced, and to take place at significantly higher temperature (as high as 10 K). Another point of disagreement is that our study yields no evidence of resilient liquid phase of the mixed clusters at temperatures lower than $\sim$ 6 K. 
Thus, the separation of the two isotopes observed at lower temperature largely reflects a process that takes place already in the high temperature liquid phase, but is then frozen in place as the temperature is lowered. While the system is eventually expected to reach its low energy (ground state) configuration, this can only happen through the slow diffusive process that takes place in an essentially crystalline matrix.

\begin{acknowledgments}
This work was supported by the Natural Science and Engineering Research Council of Canada.

\end{acknowledgments}

\section*{Conflict of interest}
The authors have no conflict of interest to disclose.

\section*{Data Availability Statement}

Raw data, including those mentioned but not shown here, as well as the computer code utilized to obtain the results illustrated herein can be obtained by contacting the authors.

\section*{References}

\bibliographystyle{aipnum4-1}
\bibliography{refs}

\end{document}